\documentclass{olplainarticle}
\usepackage{subfigure}
\usepackage{caption}
\usepackage{hyperref}

\title{Scalar Curvature in Discrete Gravity}

\author[1]{Ali H. Chamseddine}
\author[1]{Ola Malaeb}
\author[1,2]{Sara Najem}
\affil[1]{Department of Physics, American University of Beirut, Beirut, Lebanon}
\affil[2]{Center for Advanced Mathematical Sciences, American University of Beirut, Beirut, Lebanon}
\keywords{Discrete gravity, Curvature, Numerical solution.}

\begin{abstract}
We focus on studying, numerically, the scalar curvature tensor in a two-dimensional discrete space. The continuous metric of a two-sphere is transformed into that of a lattice using two possible slicings. In the first, we use two integers, while in the second we consider the case where one of the coordinates is ignorable. The numerical results of both cases are then compared with the expected values in the continuous limit as the number of cells of the lattice becomes very large.  
\end{abstract}

\begin{document}

\flushbottom
\maketitle
\thispagestyle{empty}

\section{Introduction}

Since its inception, the theory of general relativity has passed many tests. However, to date, it remains incomplete. The reason for this is attributed to inconsistencies between quantum mechanics and gravity with no unifying framework that is able to reconcile both their findings. Nevertheless, the quest for a theory that could describe the quantum behavior of the full gravitational field has been pursued along three different lines of research. The first approach, denoted by the covariant theory, quantum gravity is constructed as quantum field theory (QFT) expanding around a Minkowski background. However, firm evidence about the non-renormalizability of QFT of gravitation was found \cite{veltman}. The second line is based on canonical quantum theory in which the Hamiltonian of the classical general theory of relativity is constructed and subsequently its constraints are quantized. This approach culminated in loop quantum gravity, and despite enormous efforts during the last forty years, major problems remain. The third line includes modern methods of quantum field theory such as the covariant Feynman path integral approach, Hawking's Euclidean quantum gravity, as well as the lattice-like (discrete) alternative. Under the latter, two main discrete approaches have been mainly followed: Regge calculus \cite{regge}, and Euclidean Dynamical Triangulations \cite{EDT}. However, in both cases, progress has been very slow.

\paragraph{}
Recently, and within the lattice-like approach, a discrete gravity model was proposed by Chamseddine and Mukhanov, where the discrete analog of the Einstein action was constructed for discrete spaces \cite{discretegravity}. A space consisting of elementary cells (with Planck volume) was considered. These cells have no internal differentiable structure and are characterized by a finite number of operators and spin connections. Each cell is characterized by displacement operators that define the movement from a given cell to the next one. The diffeomorphism invariance in the continuous limit is manifested as a freedom in the choice of elementary cells. It was also shown that as the cells shrink to points, the standard formulas of differentiable geometry are recovered.\newline

\noindent Discrete gravity is promising in the sense that only a finite number of degrees of freedom is assigned to each cell for each field and this might solve the well-known problem of the ultraviolent divergences in quantum theories of gravity (check \cite{Weinberg} for example). In addition, the proposed discrete gravity avoids the problem of failure of the Liebnitz rule, which is usually a main obstacle to the development of the theory of discretized manifolds (\cite{Liebnitz}). 
\paragraph{}
In this paper, we study numerically the scalar curvature tensor in two dimensions in the discrete space. Each of the elementary cells is enumerated by two integers, $n = (n_1, n_2)$, that can take positive and negative values. These series of integers become coordinates on the manifold in the continuous limit. In the two-dimensional $(d=2)$ case that we are studying, each cell has four neighboring cells $(2d)$ which share with it a common boundary. In section two, we transform the continuous metric of a two-sphere into that of the lattice. We also explore the change in the scalar curvature in the discrete space as a function of the number of cells, and we show that it tends to the expected value in the continuous limit as the number of cells increases. Additional validation is also presented by showing that the Euler characteristic of a discretized sphere approaches the numerical value two.\newline
In section three, the values of the spin-connections and the curvature tensor in the discrete case will be compared to those obtained in the continuous case.\newline
In section four, we consider another discretization of the metric where one of the coordinates is ignorable where we show that a larger number of cells is needed to have the same accuracy as the first case.

\section{Isotropic coordinates for two-sphere}

Consider the two-sphere metric given by:
\begin{equation}
ds^{2} = a^{2}\left(  d\theta^{2}+\sin^{2}\theta d\phi^{2}\right)
\label{metric}
\end{equation}
and define:
\begin{align*}
r  &= 2\tan\frac{\theta}{2},\quad0\leq\theta\leq\frac{\pi}{2}\\
r  &  =2\cot\frac{\theta}{2},\quad\frac{\pi}{2}\leq\theta\leq\pi
\end{align*}
to get the two covering for the sphere. We then have:
\begin{align*}
ds^{2}  &  =\frac{a^{2}}{\left(  1+\frac{1}{4}r^{2}\right)  ^{2}}\left(
dr^{2}+r^{2}d\phi^{2}\right) \\
&  =\frac{a^{2}}{\left(  1+\frac{1}{4}\left(  x^{2}+y^{2}\right)  \right)
^{2}}\left(  dx^{2}+dy^{2}\right),
\end{align*}
where
\begin{equation*}
x=r\cos\phi,\quad y=r\sin\phi.
\end{equation*}
To transform this continuous metric to that on the lattice, we let:
\begin{equation*}
x=\frac{2n_{1}}{N},\quad y=\frac{2n_{2}}{N},
\end{equation*}
where
\begin{equation*}
n_{1}=0,\pm1,\pm2,\cdots,\pm\left(  N-1\right)  ,\quad n_{2}=0,\pm
1,\pm2,\cdots,\pm\left(  N-1\right),
\end{equation*}
implying that:
\begin{equation*}
0<r^{2}=\frac{4\left(  n_{1}^{2}+n_{2}^{2}\right)  }{N^{2}}\leq4
\end{equation*}
with the maximal value $r=2$ attained for $\theta=\frac{\pi}{2}.$ \newline

\begin{figure}[htbp]
     \includegraphics[scale=0.7]{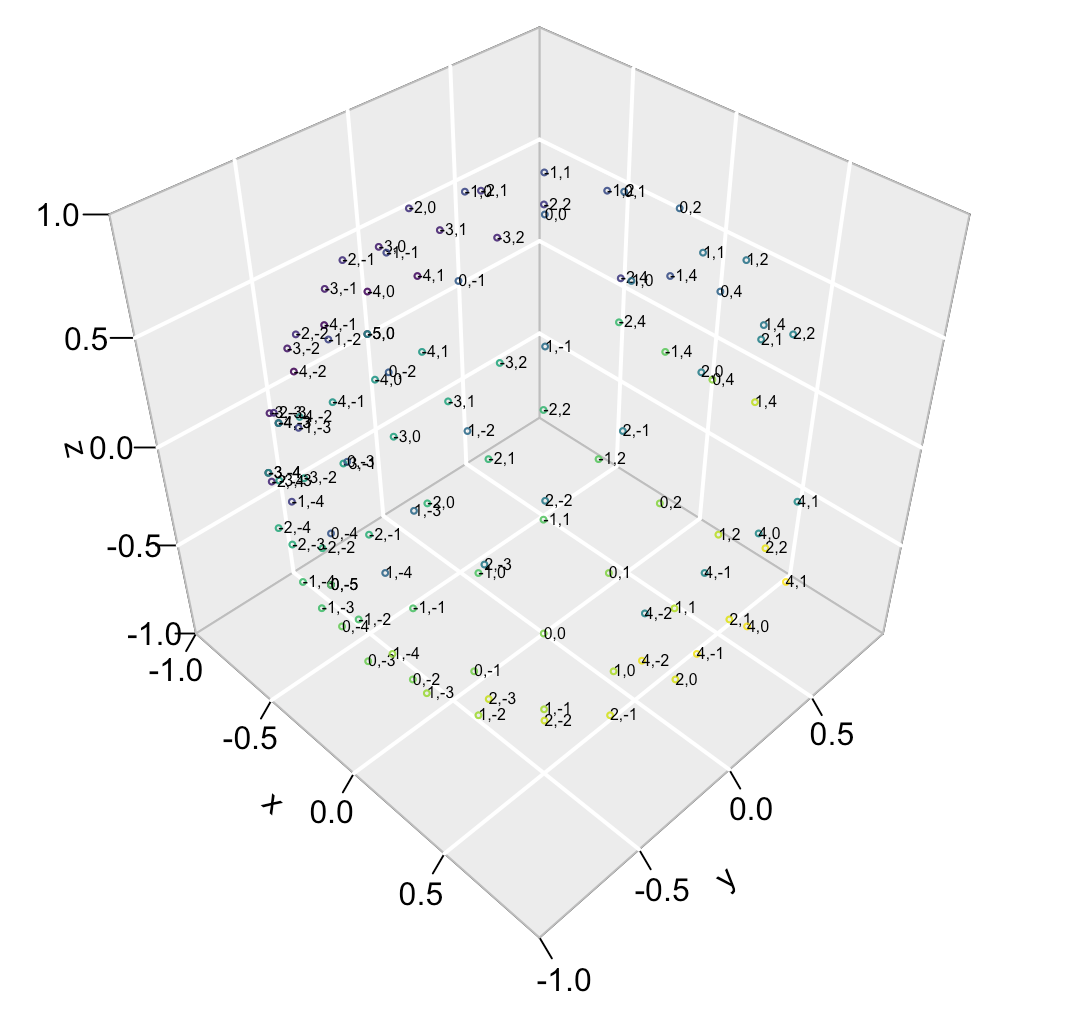}\quad  \hfill 
   \caption{Two-sphere of radius one formed from the set of the discrete points}  \label{sphere}
\end{figure}

\noindent In order to plot the set of discrete points making the discrete space, we will express the Cartesian coordinates $\left(  x_{1},x_{2},x_{3}\right)$ in terms of $n_1$ and $n_2$ to plot the discrete surface:
\begin{equation*}
x_{1}=a\sin\theta\cos\phi,\quad x_{2}=a\sin\theta\cos\phi,\quad x_{3}%
=a\cos\theta
\end{equation*}
Substituting
\begin{equation*}
x=\frac{2n_{1}}{N}=2\tan\frac{\theta}{2}\cos\phi,\quad y=\frac{2n_{2}}%
{N}=2\tan\frac{\theta}{2}\sin\phi
\end{equation*}
we can solve for the sine and cosine of $\theta$ and $\phi$. This gives finally:
\begin{equation*}
x_{1}=\frac{a}{N} \frac{2n_{1}}{1+\left(  n_{1}^{2}+n_{2}^{2}\right)  /N^{2}},\quad
x_{2}=\frac{a}{N} \frac{2n_{2}}{1+\left(  n_{1}^{2}+n_{2}^{2}\right)  /N^{2}},\quad
x_{3}=a\frac{1-\left(  n_{1}^{2}+n_{2}^{2}\right)  /N^{2}}{1+\left(  n_{1}^{2}+n_{2}^{2}\right)  /N^{2}}
\end{equation*}
which describes the upper hemisphere. We can obtain the lower hemisphere by reflection with respect to the $xy$ plane. Figure \ref{sphere} displays the set of discrete points, forming a two-sphere of radius one.

\begin{figure}[!htp]
	\centering
		{\includegraphics[scale=1]{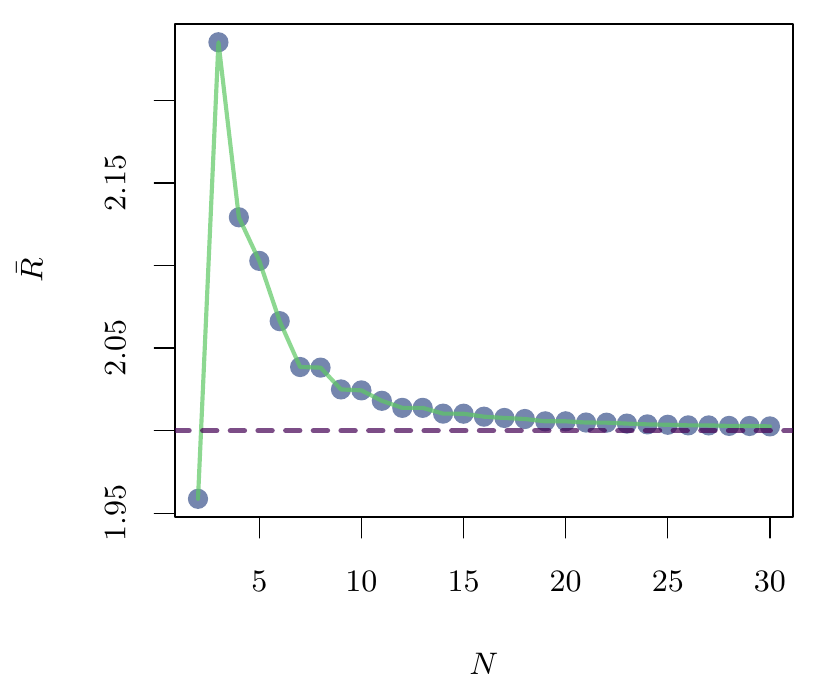}}
	    \caption{Mean of the scalar curvature versus $N$ for $a=1$} \label{curvature}
\end{figure}

\newpage
\noindent To find the expression of the scalar curvature, we start with the zweibein that are given by
\begin{equation*}
e_{1}^{1}=e_{2}^{2}=e\left(  n_{1},n_{2}\right)  =a\frac{1}{1+\frac{1}{N^{2}%
}\left(  n_{1}^{2}+n_{2}^{2}\right)  }%
\end{equation*}
Then, using equation $(54)$ in \cite{discretegravity}, we can find that $\sin\left(\frac{\pi}{4}-\omega_{1}\left(  n_{1},n_{2}\right)\right)$ is equal to
\begin{align}
&  \frac{1}{2\sqrt{2}}\left(  \left(  \frac{1}{1+\frac{1}{N^{2}}\left(
\left(  n_{1}+1\right)  ^{2}+n_{2}^{2}\right)  }\right)  ^{2}-\left(  \frac
{1}{1+\frac{1}{N^{2}}\left(  n_{1}^{2}+\left(  n_{2}+1\right)  ^{2}\right)
}\right)  ^{2}+2\left(  \frac{1}{1+\frac{1}{N^{2}}\left(  n_{1}^{2}+n_{2}%
^{2}\right)  }\right)  ^{2}\right)  \cdot\nonumber\\
&  \left(  1+\frac{1}{N^{2}}\left(  \left(  n_{1}+1\right)  ^{2}+n_{2}%
^{2}\right)  \right)  \left(  1+\frac{1}{N^{2}}\left(  n_{1}^{2}+n_{2}%
^{2}\right)  \right)
\label{omega1dis}
\end{align}
while $-\cos\left(\frac{\pi}{4}-\omega_{2}\left(  n_{1},n_{2}\right)\right)$ is equal to
\begin{align}
&  \frac{1}{2\sqrt{2}}\left(  \left(  \frac{1}{1+\frac{1}{N^{2}}\left(
n_{1}^{2}+\left(  n_{2}+1\right)  ^{2}\right)  }\right)  ^{2}-\left(  \frac
{1}{1+\frac{1}{N^{2}}\left(  \left(  n_{1}+1\right)  ^{2}+n_{2}^{2}\right)
}\right)  ^{2}+2\left(  \frac{1}{1+\frac{1}{N^{2}}\left(  n_{1}^{2}+n_{2}%
^{2}\right)  }\right)  ^{2}\right)  \cdot\nonumber\\
&  \left(  1+\frac{1}{N^{2}}\left(  n_{1}^{2}+\left(  n_{2}+1\right)
^{2}\right)  \right)  \left(  1+\frac{1}{N^{2}}\left(  n_{1}^{2}+n_{2}%
^{2}\right)  \right)
\end{align}
Using equations $(55)$ and $(56)$ appearing in \cite{discretegravity}, the curvature tensor is then given by
\begin{align}
R_{12}  =2\left(  \frac{N}{2}\right)  ^{2}\sin\left(  \frac{1}{2}\left(  \left(
\omega_{2}\left(  n_{1}+1,n_{2}\right)  -\omega_{2}\left(  n_{1},n_{2}\right)
\right)  -\left(  \omega_{1}\left(  n_{1},n_{2}+1\right)  -\omega_{1}\left(
n_{1},n_{2}\right)  \right)  \right)  \right)
\end{align}
and thus the curvature scalar
\begin{align}
R  &  =2\left(  e_{1}^{1}e_{2}^{2}\right) ^{-1} R_{12}\nonumber\\
&  =\frac{2}{a^{2}}\left(  1+\frac{1}{N^{2}}\left(  n_{1}^{2}+n_{2}%
^{2}\right)  \right)^{2} R_{12}%
\label{scalarcurvature}
\end{align}
\newline
The scalar curvature as given by equation (\ref{scalarcurvature}) was computed numerically for several values of $N$. Figure \ref{curvature} shows a plot (radius $a=1$) of the mean of the scalar curvature as $N$ varies between $2$ and $40$. As $N$ goes beyond $20$, the limit to the continuous case, $R = \frac{2}{a^2}$, is established. 

\begin{figure}[!htp]
     \includegraphics[scale=1]{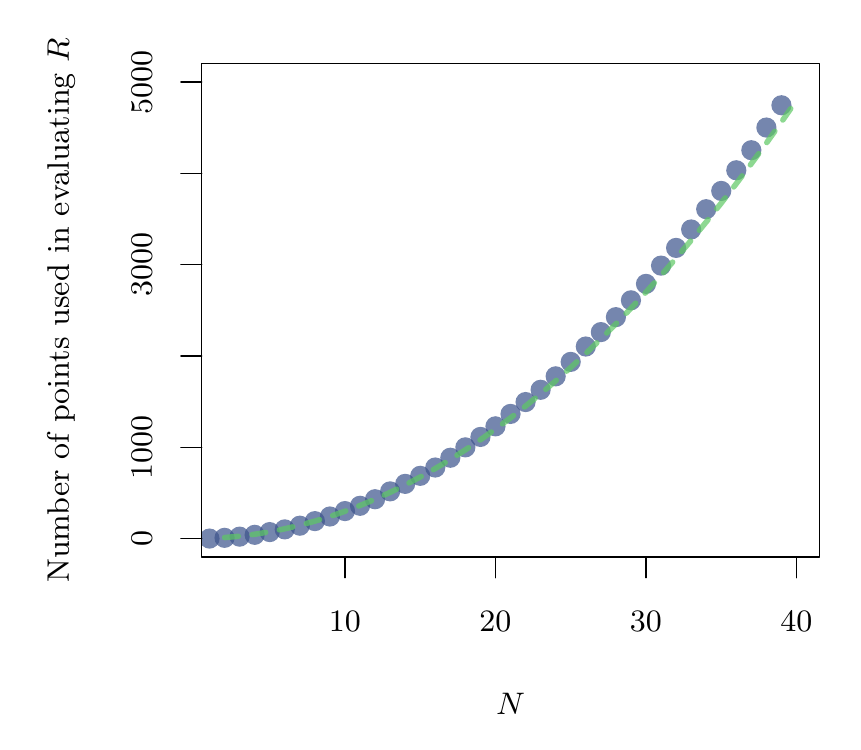}\quad  
   \caption{The estimate of the number of points used in evaluating $R$, $3N^2$, is represented by the light green curve.  The actual number of points satisfying equation (\ref{constraint}), retrieved numerically, and contributing to the computation of $R$, is represented by the blue dots.} \label{ruledout}
\end{figure}

 \newpage
Upon discretizing the space, we use $4N^2$ vertices on the lattice subject to the constraint $({n_1}^2 + {n_2}^2 \le N^2)$, where the point $(0,0)$ is at the pole, and the points satisfying $({n_1}^2 + {n_2}^2 = N^2)$ are at the equator. However, the number of points entering in the computation of the spin connections (given in terms of the zweibein at the lattice point $(n+1)$) and the curvature tensor (found in terms of $\omega(n+1)$) is constrained. The constraint is given by:
\begin{equation}
(n_1+2)^2 + {n_2}^2 < N^2 \quad \& \quad {n_1}^2 + (n_2 + 2)^2 < N^2
\label{constraint}
\end{equation} 
The constraint satisfying points as well as their estimates are shown in Figure \ref{ruledout} as a function of $N$.

\noindent The last point to be discussed in this section is the Euler characteristic \cite{euler}. This is given by \cite{gilkey}:
\begin{equation*}
\chi=\frac{1}{2\pi}{\displaystyle\int}R_{\overset{.}{1}\overset{.}{2}}^{\quad12}\left(  n\right)  dxdy
\end{equation*}
which is equal to $2$ for a sphere. To test whether this is still satisfied in the discrete case, we write the expression of $\chi$ in the discrete space. For a sphere of radius one ($a = 1$); hence, the scalar curvature is two, $\chi$ is given by:
\begin{equation*}
\chi=\frac{1}{2\pi}%
{\displaystyle\sum\limits_{n_{1}}}
{\displaystyle\sum\limits_{n_{2}}}
e^{2}\left(  n_{1},n_{2}\right)  \theta\left(  N^{2}-n_{1}^{2}-n_{2}%
^{2}\right)
\end{equation*}
where $\theta$ is the step function $\theta\left(  x\right)  =1,$ $x>0$ and $\theta\left(  x\right)  =0$ for $x<0.$ \newline

\noindent This quantity was computed numerically for $2 \le N \le 40$, and the results are outlined in the figure \ref{xi} below. 

\begin{figure}[!htp]
     \includegraphics[scale=0.8]{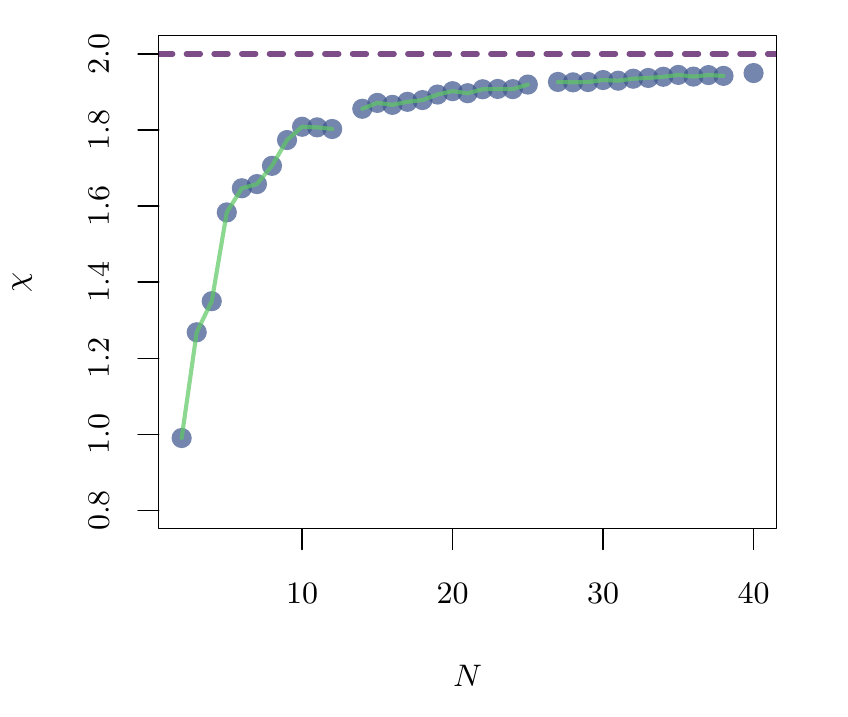}\quad  \hfill 
   \caption{The calculated Euler characteristic $\chi$ is represented by the blue dots, the light green curve is a visual guide, and the dashed line is the expected value. }  \label{xi}
\end{figure}


\section{Continuous case}

In this section, we will look at the expressions of the spin connections and curvature tensor in the continuous case and compare with those got in the discrete space. For the continuous case, we have
\begin{align}
\omega_{\overset{.}{1}}^{12}  &  \equiv\omega_{\overset{.}{1}}=-\frac
{y}{2\left(  1+\frac{1}{4}r^{2}\right)  } \label{omega1cont} \\
\omega_{\overset{.}{2}}^{12}  &  \equiv\omega_{\overset{.}{2}}=\frac
{x}{2\left(  1+\frac{1}{4}r^{2}\right)  }\\
R_{\overset{.}{1}\overset{.}{2}}^{12}  &  =-\frac{1}{\left(  1+\frac{1}%
{4}r^{2}\right)  ^{2}}
\end{align}
The north pole is at the point $\theta=0$ for which $x=y=0,$ so that
$\omega_{\overset{.}{1}}=0=\omega_{\overset{.}{2}}$. The metric is not
singular at this point. However, the south pole corresponds to the point
$\theta=\pi$ for which $r\rightarrow\infty$ and the metric becomes singular,
$\omega_{\overset{.}{1}}$ and $\omega_{\overset{.}{2}}\rightarrow0$ but the
curvature is not singular, showing that this is a coordinate singularity. This
is the reason that for $\theta>\frac{\pi}{2}$ we have to use the variable
$r=2\cot\frac{\theta}{2}.$

\begin{figure}[!htp]
     \includegraphics[scale=0.88]{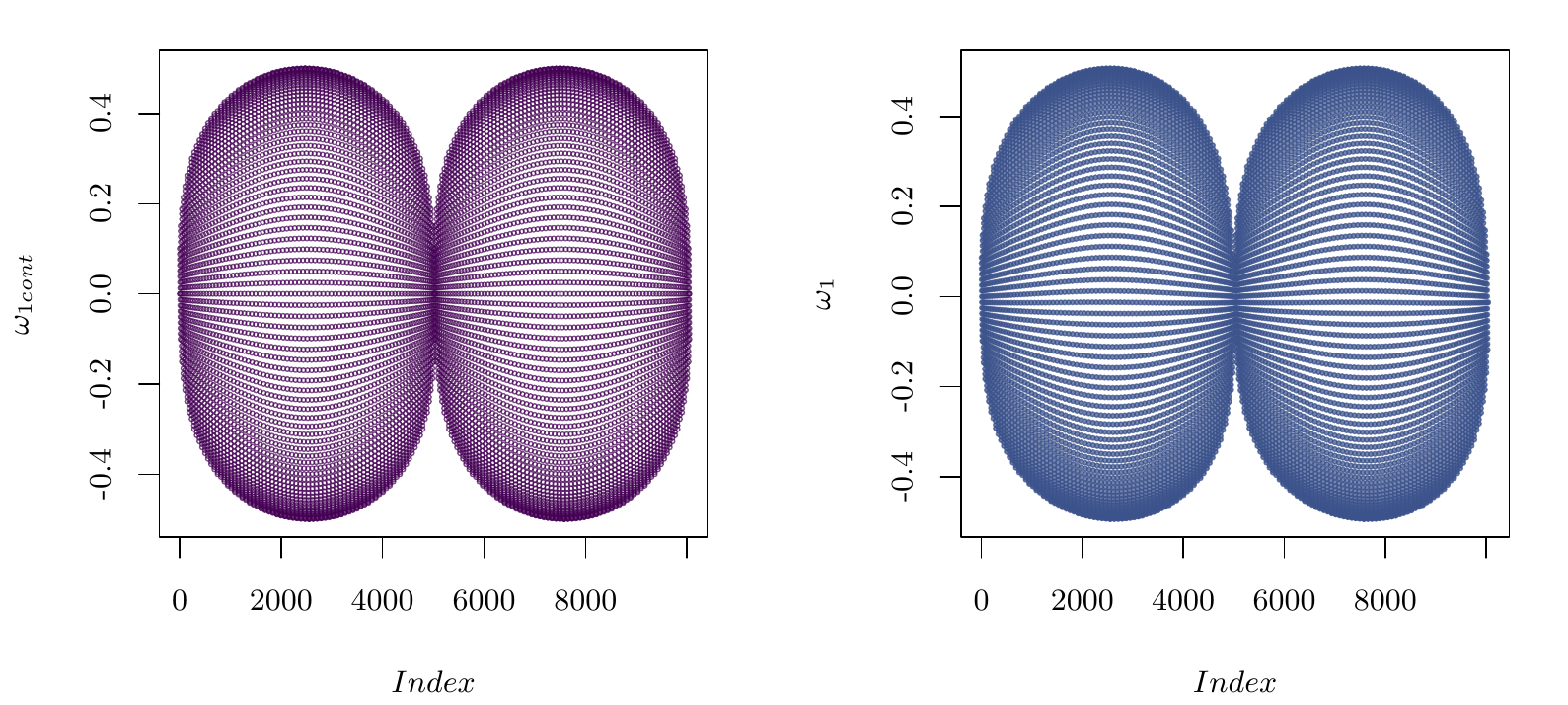}\quad\hfill 
   \caption{The continuous limit of the curvature $\omega_{1cont}$ is shown in the left, while its discrete counterpart is shown on the right for $N=40$.}  \label{omegacont}
\end{figure}

\begin{figure}[!htp]
     \includegraphics[scale=0.8]{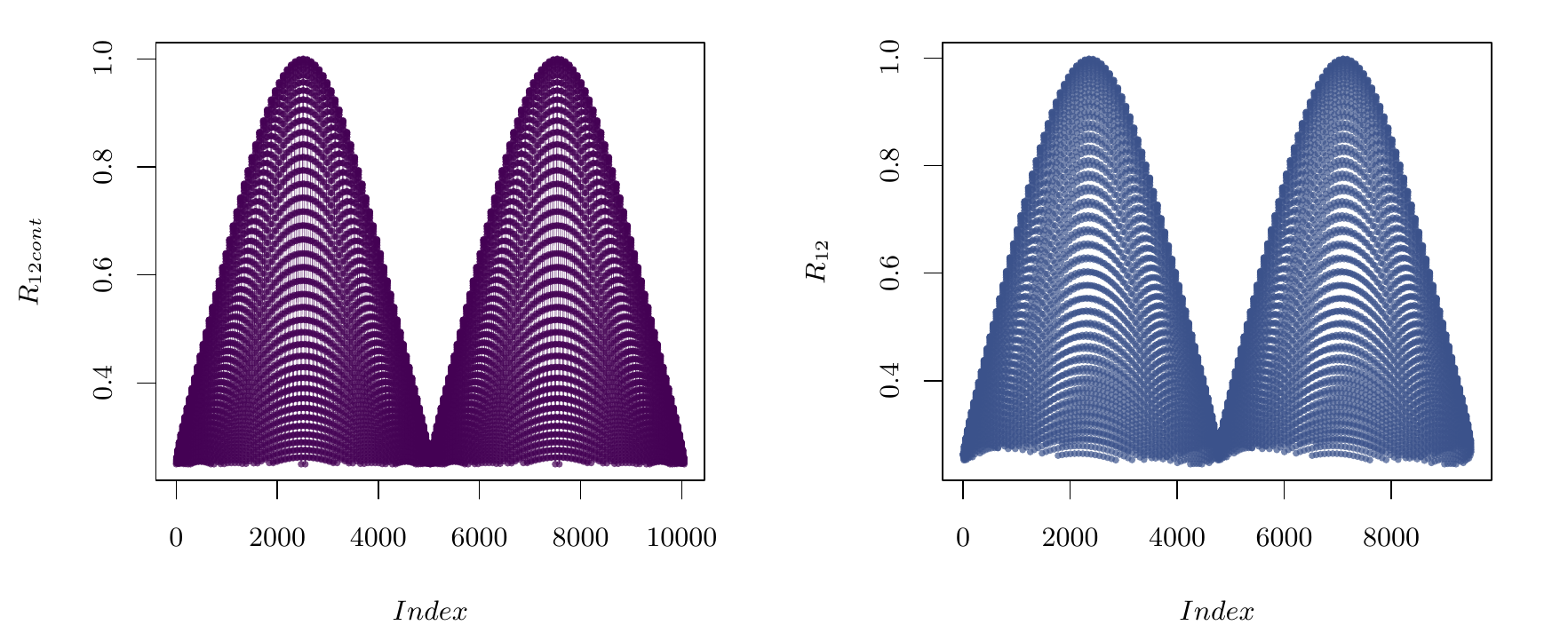}\quad  \hfill 
   \caption{The continuous limit of the $R_{12 \ cont}$ is shown in the left, while its discrete counterpart is shown on the right for $N=40$.}  \label{r12cont}
\end{figure}

\noindent Figure \ref{omegacont} shows the plot of $\omega_1$ as computed using equation (\ref{omega1cont}), on the left, compared to that of $\omega_1$ in the discrete space, on the right, as given by equation (\ref{omega1dis}). In a similar manner, Figure \ref{r12cont} compares the curvature tensor in the continuous and that in discrete space on the left and right figures respectively. Both plots are done for $N=40$. This shows the that the continuous limit is recovered from the discrete case as more lattice points are included in the computations, which corresponds to finer discretization.  

\section{Ignorable coordinates}

\begin{figure}[!htp]
    \centering
     \includegraphics[scale=0.8]{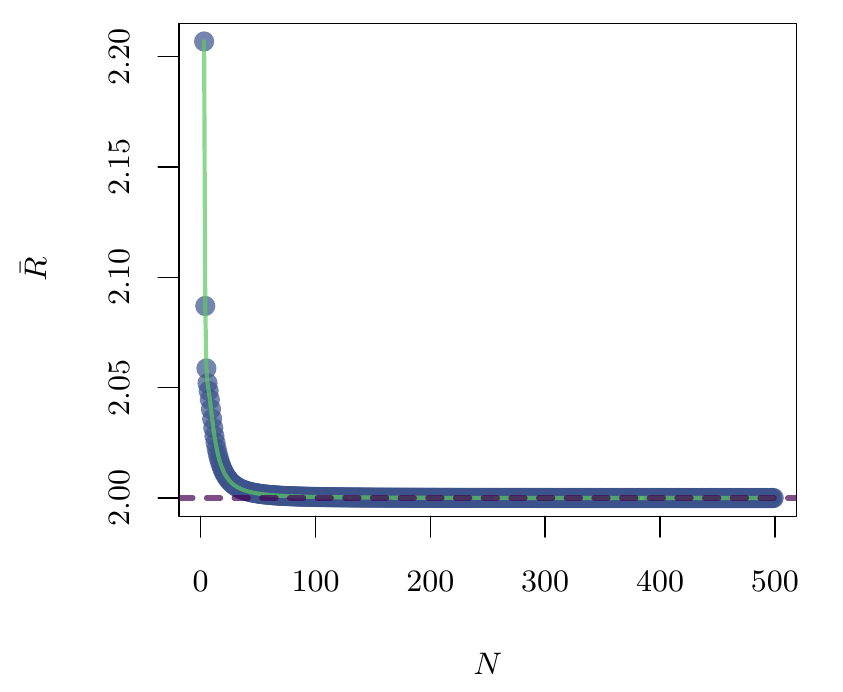}\quad  \hfill 
   \caption{The scalar curvature is followed as a function of $N$ for the ignorable coordinate case for $a = 1$, with the dashed lines representing the expected value at 2.}  \label{ignorableR}
\end{figure}

Another way of discretizing the 2-d sphere metric which is given by equation \ref{metric} is by setting
\begin{align*}
    \theta =\frac{n\pi}{N} \quad \& \quad \phi=\frac{2m\pi}{M}
\end{align*}
where $n=0,1,\cdots N-1$ and $m=0,1,\cdots M-1$. For simplicity we take $M=2N$ so that
$\Delta\theta=\frac{\pi}{N}=\Delta\phi.$ We also replace the function
$\sin\theta$ with $SIN[\frac{n\pi}{N}]$ defined by:
\begin{equation*}
SIN[\frac{n\pi}{N}]=\frac{1}{2i}\left(  \left(  1+\frac{i\pi}{N}\right)
^{n}-\left(  1-\frac{i\pi}{N}\right)  ^{n}\right)
\end{equation*}
satisfying the relation below:
\begin{align*}
\Delta_{\theta}SIN[\frac{n\pi}{N}]  &  \equiv\frac{1}{\Delta\theta}\left(
SIN[\frac{\left(  n+1\right)  \pi}{N}]-SIN[\frac{n\pi}{N}]\right) \\
&  =COS[\frac{n\pi}{N}],
\end{align*}
where
\begin{equation*}
COS[\frac{n\pi}{N}]=\frac{1}{2}\left(  \left(  1+\frac{i\pi}{N}\right)
^{n}+\left(  1-\frac{i}{N}\pi\right)  ^{n}\right).
\end{equation*}
The vierbein components are:
\begin{equation*}
e_{\overset{.}{1}}^{1}\left(  n_{1},n_{2}\right)  =a,\quad e_{\overset{.}{2}%
}^{2}\left(  n_{1},n_{2}\right)  =ae\left(  n\right).
\end{equation*}
Thus $n_{2}$ will be ignorable coordinate and for simplicity we will set $n_{1}=n.$  In this case, the spin connections are given by:
\begin{equation*}
\sin\omega_{1}=\frac{-\frac{1}{2e\left(  n+1\right)  } + e\left(  n\right)
\sqrt{e^{2}\left(  n\right)  +1-\frac{1}{4e^{2}\left(  n+1\right)  }}}%
{1+e^{2}\left(  n\right)  }%
\end{equation*}
In this expression $\omega_{1}$ stands for $\frac{\pi}{N}\omega_{1}$ and in the limit of very large $N$ we have $\omega_{1}\rightarrow0$. For $\omega_{2}$ we get:
\begin{equation*}
\sin\omega_{2}=\frac{e\left(  n\right)  \left(  1+\frac{e^{2}\left(  n\right)
-e^{2}\left(  n+\right)  }{2}\right)  - \sqrt{e^{2}\left(  n+1\right)
+\frac{1}{2}e^{2}\left(  n+1\right)  e^{2}\left(  n\right)  -\frac{1}{4}%
e^{4}\left(  n+1\right)  -\frac{1}{4}e^{4}\left(  n\right)  }}{1+e^{2}\left(
n\right)  }%
\end{equation*}
Again $\omega_{2}$ stands for $\frac{\pi}{N}\omega_{2}$, and we know that in
the limit of large $N$, we have the limit $\omega_{2}\rightarrow\frac{N}{\pi
}\cos\frac{n\pi}{N}$. The curvature tensor is then:
\begin{align*}
R_{\overset{.}{1}\overset{.}{2}}^{\quad}\left(  n\right)   &  =\frac{2}%
{\Delta\theta\Delta\phi}\sin\left(  \frac{1}{2}\left(  \omega_{\overset{.}{2}%
}\left(  n+1\right)  -\omega_{\overset{.}{2}}\left(  n\right)  \right)
\right) \\
&  =2\left(  \frac{N}{\pi}\right)  ^{2}\sin\left(  \frac{1}{2}\left(
\omega_{\overset{.}{2}}\left(  n+1\right)  -\omega_{\overset{.}{2}}\left(
n\right)  \right)  \right)
\end{align*}
and the curvature scalar is given by:
\begin{equation*}
R=\frac{4}{e\left(  n\right)  }\left(  \frac{N}{\pi a}\right)  ^{2}\sin\left(
\frac{1}{2}\left(  \omega_{\overset{.}{2}}\left(  n+1\right)  -\omega
_{\overset{.}{2}}\left(  n\right)  \right)  \right).
\end{equation*}
Figure \ref{ignorableR} displays a plot, for $a = 1$, of the scalar curvature versus $N$. Again, this reveals that the expected value in the continuous limit is obtained. 

\newpage
\noindent Notice that a much larger value of $N$ corresponding to much finer lattice is needed in this case to achieve the same accuracy achieved before with non-ignorable coordinates. In the latter case, $N$ represents the number of discrete points, whereas in the former the number of discrete points is $\propto N^2$. Because the angle $\phi$ is ignorable, for every angle $\theta=\frac{n\pi}{N}$ we will have all the points with angle $\phi=\frac{m2\pi}{M}$ on the circle with radius $SIN[n].$ For small values of $N$ the shape will be that of ellipsoid, but for larger $N$ it will approach a sphere. For example, the area
of the surface is:
\begin{equation}
{\displaystyle\sum\limits_{n=0}^{N-1}}
{\displaystyle\sum\limits_{m=0}^{M-1}}
\left(  \frac{\pi}{N}a\right)  ^{2}SIN[n]=\frac{\pi^{2}a^{2}}{Ni}
{\displaystyle\sum\limits_{n=0}^{N-1}}
\left(  \left(  1+\frac{i\pi}{N}\right)  ^{n}-\left(  1-\frac{i\pi}{N}\right)
^{n}\right)
\label{area}
\end{equation}

\noindent Figure \ref{ignorable} below shows a plot (radius $a=1$) for the area, computed numerically using equation (\ref{area}), as $N$ varies between $2$ and $500$. As $N$ goes beyond $100$, the shape is approaching that of a sphere, with an area of $4 \pi a^2$.

\begin{figure}[!htp]
  \centering
     \includegraphics[scale=0.8]{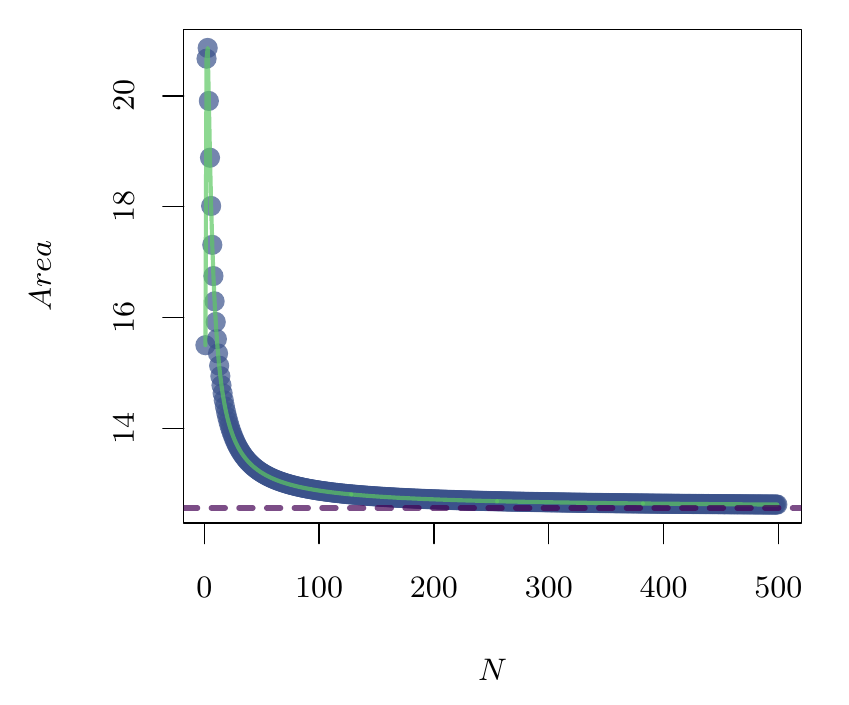}\quad  \hfill 
   \caption{The area is followed as a function of $N$ for the ignorable coordinate case for $a = 1$.}  \label{ignorable}
\end{figure}

\clearpage

\section{Conclusions}

In this paper, we discussed the recently proposed model of discrete gravity where each elementary cell is characterized by displacement operators that connect a cell to a neighboring one. Using the definition of the curvature of the discrete space, the curvature of a two-dimensional sphere was computed numerically in two cases: isotropic coordinates and ignorable coordinates. It was shown that as the cells shrink to points, the standard curvature for a sphere in the continuous manifold was recovered. Our results show that the proposal of Chamseddine and Mukhanov is very effective in computing the curvature of discrete spaces. Our next step is to perform a numerical study of the discretization of three- and four-dimensional manifolds. 

\section*{Acknowledgments}
A.H.C would like to thank Slava Mukhanov for suggesting the use of isotropic coordinates. The work of A. H. C is supported in part by the National Science Foundation Grant No. Phys-1912998. 
\section*{Supporting Material}
The scripts used in the paper are available at the following 
 \href{https://www.dropbox.com/sh/yf016aozwgr1kgj/AACe6gJyFyr9OuoS_mjD-z_ta?dl=0}{link.}

\bibliography{references}

\end{document}